\documentclass[prd,nofootinbib,floats,superscriptaddress,eqsecnum,tightenlines,preprintnumbers,11pt]{revtex4}

\usepackage{hyperref}
\usepackage{graphicx}
\usepackage{amsmath,amssymb,amsfonts,amsthm,latexsym,mathtools}
\usepackage{csquotes}
\usepackage{xcolor}
\usepackage{physics}
\usepackage{bm}
\usepackage{subcaption}
\usepackage{tensor}
\usepackage{array}
\usepackage{siunitx}

\usepackage{calc}

\newcommand{\bb}{f_1}
\newcommand{\gp}{{p_1}}
\newcommand{\rz}{r_-}
\newcommand{\rp}{r_+}

\begin{document}

\title{Numerical computation of quasinormal modes \\ in \\ 
the first-order approach to black hole perturbations in  modified gravity}

\author{Hugo Roussille}
\affiliation{Univ Lyon, ENS de Lyon, CNRS, Laboratoire de Physique, F-69342 Lyon, France}
\author{David Langlois}
\affiliation{Laboratoire Astroparticule et Cosmologie, CNRS, Universit\'e Paris Diderot Paris 7, 75013 Paris, France}
\author{Karim Noui}
\affiliation{Laboratoire de Physique des deux Infinis IJCLab, Universit\'e Paris Saclay, France}
\affiliation{Laboratoire Astroparticule et Cosmologie, CNRS, Universit\'e Paris Diderot Paris 7, 75013 Paris, France}

\date{\today}

\begin{abstract}
We present a novel approach to the numerical computation of quasi-normal modes, based on the first-order (in radial derivative) formulation of the equations of motion and using a matrix version of the continued fraction method. This numerical method is particularly suited to the study of static black holes in modified gravity, where the traditional second-order, Schr\"odinger-like, form of the equations of motion is not always available. Our approach relies on the knowledge of the asymptotic behaviours of the perturbations near the black hole horizon and at spatial infinity, which can be obtained via the systematic algorithm that we have proposed recently.
%Recently, it has been proposed a method to study perturbations about black holes (and more generally static and spherically symmetric compact objects) in scalar-tensor theories 
%which relies on a first order formulation of the perturbations equations. In particular, it has been shown how to extract the asymptotic behaviour of the perturbations at the horizon and at spatial infinity, which is necessary  to define  quasi-normal modes (QNMs).  
In this work, we first present our method for the perturbations of a Schwarzschild black hole and show that we recover the well-know frequencies of the QNMs to a very high precision. We then apply our method to 
%In this work, we go one step further and we present an adaptation of the continued fraction method that allows us to compute precisely QNMs directly from such a first-order system. As a proof of concept, we apply it to
the axial perturbations of an exact black hole solution in a particular scalar-tensor theory of gravity.  We  also cross-check the obtained QNM frequencies with other numerical methods.
\end{abstract}

\maketitle

%\tableofcontents

\section{Introduction}

The recent detection of gravitational waves (GW) has opened a new window on  gravitational physics  by giving unprecedented access to the regime of strong gravity.
So far, the GW  measurements 
are in good agreement with the predictions of general relativity (GR), but the 
rapidly increasing
 number of events and the improved sensitivity of the detectors expected in the near future  will enable to verify  GR to a  high degree of precision or, alternatively, to detect some deviations from GR.

In this perspective, it is important to consider  alternative theories of gravity, or extensions of GR, and study how their predictions deviate from GR, so that future analyses can extract the most relevant information from   upcoming GW data. Most extensions of GR are scalar-tensor theories, which involve, directly or in disguise,  a scalar field in addition to the usual metric tensor. The most general scalar-tensor theories that propagate a single scalar degree of freedom  have been classified within the framework of DHOST theories, allowing for higher-order derivatives of the scalar field in the action \cite{Langlois:2015cwa,Langlois:2015skt,BenAchour:2016cay,Crisostomi:2016czh,BenAchour:2016fzp} (see \cite{Langlois:2018dxi,Kobayashi:2019hrl} for reviews).  These theories  
possess
 a very rich phenomenology and  admit a  number of exact static and spherically symmetric solutions (black holes or more exotic objects) even though very few exact rotating solutions are known \cite{Babichev:2017guv,BenAchour:2018dap,Motohashi:2019sen,Charmousis:2019vnf,Minamitsuji:2019shy,BenAchour:2020wiw,Minamitsuji:2019tet,Anson:2020trg, BenAchour:2020fgy,Takahashi:2020hso,Babichev:2020qpr,Baake:2020tgk,Capuano:2023yyh,Bakopoulos:2023fmv} (see also the review \cite{Babichev:2023psy}).

In the GW signal from a binary black hole merger,  the ringdown phase is particularly interesting as it can be described  by linear perturbations about a stationary black hole (BH) and is therefore easier to predict in theories of modified gravity than the whole inspiral phase. The ringdown signal  mainly corresponds to a superposition of quasinormal modes, whose frequencies are quantised (see e.g.  \cite{Kokkotas:1999bd, Nollert:1999ji, Berti:2009kk, Konoplya:2011qq} for reviews) and its detailed analysis via so-called \enquote{black hole spectroscopy}, represents an invaluable tool to test GR and  look for characteristic signatures of modified gravity \cite{Berti:2005ys,Berti:2018vdi}. 

In parallel to semi-analytical methods (see e.g. \cite{Hui:2022vov} for recent works), many numerical  techniques have been developed to compute QNMs \cite{Berti:2009kk,Pani:2013pma,Franchini:2023eda}. One particularly efficient method was introduced  by Leaver \cite{Leaver:1985ax}, who managed  to compute a large number of QNMs for the Schwarzschild and Kerr solutions with a high level of accuracy. His idea was to start from the equation of motion for the BH perturbations,  written in the form of a second-order 
Schr\"odinger-like equation,  thanks to the seminal works of  Regge-Wheeler \cite{Regge:1957td} and Zerilli \cite{Zerilli:1970se} for Schwarzschild, and of Teukolsky \cite{Teukolsky:1973ha} for Kerr, and to construct an ansatz for the solution that automatically implements the appropriate boundary conditions near the BH horizon and at spatial infinity. The ansatz depends on a bounded function, which can be expanded in a power series, and the equation of motion translates 
 into a three-term recurrence relation for the coefficients of the series. A viable solution corresponds to a convergent series, associated  with the so-called minimal solution of the recurrence relation, which can be obtained by solving a continued-fraction equation.
 
 In theories of  modified gravity, the equations of motion for the BH perturbations become more complicated than in GR and a nice second-order 
Schr\"odinger-like equation is  not always available. This is why we have recently developed a new approach that directly extracts the asymptotic behaviour of the perturbations from the first-order system of the equations of motion \cite{Langlois:2021xzq}.  In the present work, we use this approach to construct an ansatz for the solution of the first-order system, with the appropriate asymptotic behaviour. The ansatz now depends on {\it several} bounded functions which, when expressed in power series,  must satisfy  a {\it matrix} recurrence relation, which we solve numerically by using a matricial version of the continued fraction method. 

In order to test this new numerical technique we have first applied it to the familiar Schwarzschild case, starting directly from the first-order system of equations, in contrast with Leaver's approach. We show that the well-known QNM frequencies for Schwarzschild can be recovered in this way, with a high precision.  We then consider  the axial perturbations of an exact BH solution in a scalar-tensor theory, constructed in \cite{Babichev:2017guv} and 
{dubbed BCL here}, which can be seen as a one-parameter deformation of Schwarzschild. Using our  first-order approach, we compute numerically the QNM frequencies of the axial perturbations, for different values of the parameter. We also cross-check the robustness of our technique with different numerical tests and comparison with other methods. While the first-order system we study is of dimension 2, our method can in principle be applied to higher-dimensional systems such as the ones appearing in the polar sector of perturbations \cite{Langlois:2021aji,Langlois:2022eta}. Note that a first order approach has already been used,  in a case where the asymptotic analysis of the perturbations is straightforward, to compute numerically QNMs with
a  shooting method  (e.g. in \cite{Blazquez-Salcedo:2016enn,Blazquez-Salcedo:2017txk} for Einstein-Gauss-Bonnet-Dilaton gravity).  

This paper is structured as follows. In the next section, after a brief review of the derivation of the equations of motion in the Schwarzschild case, we explain in detail the method of matrix continued fraction for solving the first-order system of  equations and show that we recover the standard values of the Schwarzschild QNMs. In the subsequent section, 
we present the exact BH in modified gravity  and give the first-order system of equations satisfied by the axial perturbations. In section \ref{sec:QNMs-BCL}, we apply our numerical technique to this new system and obtain the QNMs, which depend on a single parameter, thus providing a continuum of modes in the complex plane relating  the Schwarzschild BH QNMs and those of this family of solutions. In section IV, we present numerical convergence and consistency checks. We finally conclude with some perspectives. We also provide some details  and comparisons with other numerical methods  in the appendices.

\section{Schwarzschild BH QNM spectrum}
In this section, we 
briefly recall  how to derive the equations for linear perturbations about a Schwarzschild BH in GR, obtaining directly
 a system of two coupled  first-order  equations.  Then, we use this ``simple'' example to illustrate how one can adapt  the well-known continuous fraction method (originally used to compute the spectrum of Schr\"odinger-like operators) to such a first order matrix system.

\subsection{Dynamics of linear perturbations}
\label{sec:GR-perturbations} 

\subsubsection{Perturbation equations in a first-order system }
To derive the equations of motion for the linear perturbations, we substitute in  the GR action,
\begin{equation}
	S_\mathrm{GR}[g_{\mu\nu}] = \int\dd[4]{x} \sqrt{-g} \, R \,,
	\label{eq:action-GR}
\end{equation}
the perturbed metric 
\begin{equation}
	g_{\mu\nu} = \bar{g}_{\mu\nu} + h_{\mu\nu} \,,
\end{equation}
where  $\bar{g}_{\mu\nu}$ is the background solution and  $h_{\mu\nu}$   the perturbation, and expand the action up to  quadratic order in $h_{\mu\nu}$. 
The dynamics of the perturbations is governed by 
%we obtain
%
%\begin{equation}
%	S_\mathrm{GR}[g_{\mu\nu}] = S_\mathrm{GR}[\bar{g}_{\mu\nu}] + S_\mathrm{quad}[h_{\mu\nu}]\,,
%\end{equation}
%
 the quadratic action $S_\mathrm{quad}$, which  reads, when the Ricci tensor vanishes {(since we are in vacuum)},
\begin{align}
	S_\mathrm{quad}[h_{\mu\nu}] = \int\dd[4]{x} \sqrt{-\bar{g}} \Big[
	%&-\frac12 \tensor{h}{_\mu_\nu}\tensor{h}{^\mu^\nu} \bar R
	%+ \frac14 h^2 \bar  R
	%+ h \tensor{h}{_\mu_\nu} \tensor{\bar R}{^\mu^\nu}
	%+ 4 \tensor{h}{_\mu^\rho} \tensor{h}{^\mu^\nu} \tensor{\bar R}{_\nu_\rho} 
	%- 2 \tensor{h}{^\mu^\nu} \tensor{h}{^\rho^\sigma} \tensor{\bar R}{_\mu_\rho_\nu_\sigma}
	%\nonumber\\
	& \quad \frac12 (\bar\nabla_\mu h) (\bar\nabla^\mu h)
	- 2 (\bar\nabla_\mu \tensor{h}{^\mu_\nu}) (\bar\nabla_\rho \tensor{h}{_\nu^\rho})
	- (\bar\nabla_\mu h)(\bar\nabla_\nu \tensor{h}{^\mu^\nu})
	\nonumber\\
	&\qquad+ 3 (\bar\nabla_\nu \tensor{h}{_\mu_\rho}) (\bar\nabla^\rho \tensor{h}{^\mu^\nu})
	- \frac12 (\bar\nabla_\rho \tensor{h}{_\mu_\nu}) (\bar\nabla^\rho \tensor{h}{^\mu^\nu})
	- 2 \tensor{\bar R}{_\mu_\rho_\nu_\sigma} \tensor{h}{^\mu^\nu} \tensor{h}{^\rho^\sigma} 
	\Big] \,,
\end{align}
where $\bar{R}_{\mu\nu\rho\sigma}$ is the Riemann tensor of  the background metric $\bar{g}_{\mu\nu}$ and $\bar{\nabla}$ denotes the covariant derivative compatible with $\bar{g}_{\mu\nu}$. 

For a static and spherically symmetric background, the metric is of the form 
\begin{equation}
\bar{g}_{\mu\nu} \dd{x^\mu} \dd{x^\nu} = - A(r) \dd{t}^2 + \frac{\dd{r}^2}{B(r)} + C(r)\dd{\Omega}^2 \,,\nonumber\\
\end{equation}
and in the particular case of the Schwarzschild solution, we have
\begin{equation}
A(r) = B(r) = 1 - \frac{\mu}{r} \,,\quad C(r) = r^2 \,,
\end{equation}
where $\mu$ 
is a  constant corresponding to twice the black hole mass. 

The perturbations  $h_{\mu\nu}$ are conveniently described 
via 
a $2+2$ decomposition onto the sphere, in which the various  components are expanded in  spherical harmonics $Y_{\ell m}(\theta, \varphi)$ according to
\begin{align}
	h_{tt} &= A(r) \sum_{\ell,\,m} H_0^{\ell m}(t,r)  Y_{\ell m}(\theta, \varphi) \,,\nonumber  \\ h_{tr} &= \sum_{\ell,\,m} H_1^{\ell m}(t,r)  Y_{\ell m}(\theta, \varphi) \,,\nonumber\\
	h_{rr} &= \frac{1}{B(r)} \sum_{\ell,\,m} H_2^{\ell m}(t,r)  Y_{\ell m}(\theta, \varphi) \,, && \nonumber\\
	h_{ta} &= \sum_{\ell,\,m} \qty[\beta^{\ell }(t,r) \partial_a + h_0^{\ell m}(t,r) \sin\theta \, \epsilon_{ab}\partial^b]Y_{\ell m}(\theta, \varphi) \,,\nonumber \\
	 h_{ra} &= \sum_{\ell,\,m} \qty[\alpha^{\ell m}(t,r)\partial_a + h_1^{\ell m}(t,r) \sin\theta \, \epsilon_{ab} \partial^b] Y_{\ell m}(\theta, \varphi) \,,\nonumber\\
	\omit\rlap{$\displaystyle h_{ab} = \frac12 \sum_{\ell,\,m}   h_2^{\ell m}(t,r)\sin\theta ( \epsilon_{ac}\nabla^c\nabla_b + \epsilon_{bc}\nabla^c\nabla_a) Y_{\ell m}(\theta, \varphi) \, $}  \nonumber \\
	& +  \sum_{\ell,\,m} \qty[K^{\ell m}(t,r) g_{ab} + G^{\ell m}(t,r) \nabla_a\nabla_b]Y_{\ell m}(\theta, \varphi)  \, ,
	\label{eq:RW-gauge}
\end{align}
where $h_0$, $h_1$, $h_2$, $H_0$, $H_1$, $H_2$, $\alpha$, $\beta$, $K$ and $G$ are  functions of $(t,r)$. The indices $a$ and $b$ denote the angular coordinates
 $\{\theta, \varphi\}$ and   $\epsilon_{ab}$  is the fully antisymmetric tensor associated with the metric of the 2-sphere.  Since there is no mixing between modes with different values of $\ell$ and $m$ at the linear level, we will drop these labels in the following. Furthermore, it is convenient to work in the frequency domain so that any function $f(t, r)$ is replaced by  ${f}(r) e^{-i\omega t}$.

The equations of motion for the perturbations are given by
\begin{equation}
	\mathcal{E}_{\mu\nu} \vcentcolon= \fdv{S_\mathrm{quad}}{h_{\mu\nu}} = 0 \,.
	\label{eq:pert-eqs}
\end{equation}
By a gauge transformation, one can always take $\alpha$, $\beta$, $G$ and $h_2$ to be zero \cite{Kobayashi:2012kh, Kobayashi:2014wsa}: this is the well-known Regge-Wheeler gauge \cite{Regge:1957td}.  Substituting the expressions~\eqref{eq:RW-gauge} into the perturbation equations~\eqref{eq:pert-eqs} yields two decoupled systems (see~\cite{Langlois:2021xzq}) which can be written in a matrix form as follows,
\begin{equation}
	\dv{X_\mathrm{ax}}{r} = M_\mathrm{ax} X_\mathrm{ax} \, , \qquad \dv{X_\mathrm{po}}{r} = M_\mathrm{pol} X_\mathrm{pol} \,.
\end{equation}
The first system corresponds to axial perturbations of the Schwarzschild black hole with
\begin{equation}
	X_\mathrm{ax} = \begin{pmatrix} h_0 \\ h_1 \end{pmatrix} \qq{and} M_\mathrm{ax} = \begin{pmatrix}
		\frac{2}{r} & -i\omega + \frac{2i\lambda(r-\mu)}{\omega r^3} \\
		-\frac{i\omega r^2}{(r-\mu)^2} & -\frac{\mu}{r(r-\mu)}
	\end{pmatrix} \,,
	\label{eq:first-order-schwarzschild-axial}
\end{equation}
where 
\begin{equation}
	\lambda = \frac{\ell(\ell + 1)}{2} - 1 \,.
\end{equation}
%
%{\bf [Add expression in terms of $A$, $B$ and $C$ for the axial sector ?]}
The second system corresponds to polar perturbations with
\begin{eqnarray}
	&&X_\mathrm{pol} = \begin{pmatrix} K \\ \flatfrac{H_1}{\omega} \end{pmatrix} \quad \text{and} \nonumber \\
	&&M_\mathrm{pol} = \frac{1}{3\mu + 2\lambda r} \begin{pmatrix}
		\frac{\mu  (3 \mu +(\lambda -2) r) - 2 r^4 \omega ^2}{r (r-\mu ) } & \frac{2 i (\lambda +1) (\mu
			+\lambda  r)+2 i r^3 \omega ^2}{r^2 } \\
		\frac{i r \left(9 \mu ^2-8 \lambda  r^2+8 (\lambda -1) \mu  r\right) + 4 i r^5 \omega ^2 }{2 (r-\mu )^2 } & \frac{2 r^4 \omega ^2-\mu  (3 \mu +3 \lambda  r+r)}{r (r-\mu )} \\
	\end{pmatrix} \,.
\end{eqnarray}
In the following, we focus our attention on the system of axial perturbations. 

%Our aim is to illustrate how one can adapt the continuous fraction method (which has mainly being used from the Schrödinger-like formulation of the perturbations) to compute QNMs directly from the first order system~\eqref{eq:first-order-schwarzschild-axial}.

\subsubsection{Definition of quasinormal modes}
\label{sec:def-QNMs}

As described in \cite{Langlois:2021xzq}, one can extract the asymptotic behaviours of the perturbations, at the horizon and at spatial infinity,   directly from the first-order system~\eqref{eq:first-order-schwarzschild-axial}  by resorting to  an asymptotic expansion of the matrix $ M_\mathrm{ax}$.  

Introducing the traditional tortoise coordinate defined by
\begin{equation}
\label{tortoise}
	\dv{r_*}{r} = \frac{1}{1 - \flatfrac{\mu}{r}} \, \Longrightarrow \; r_* = r + \mu \ln(r-\mu) \,,
\end{equation}
the leading-order  terms in the asymptotic expansion of the perturbations at the horizon (i.e. when $r \to \mu$ or equivalently $r_* \to -\infty$) were found to be given by \cite{Langlois:2021xzq}
\begin{align}
	h_0(r) &= (c^{\mathrm{hor}}_+ e^{i\omega r_*} + c^{\mathrm{hor}}_- e^{-i\omega r_*}) (1 + \mathcal{O}(r-\mu)) \,,\nonumber\\
	h_1(r) &= \frac{\mu}{r-\mu} (-c^{\mathrm{hor}}_+ e^{i\omega r_*} + c^{\mathrm{hor}}_- e^{-i\omega r_*}) (1 + \mathcal{O}(r-\mu)) \,,
	\label{eq:asymp-schwa-horiz}
\end{align}
where $c^{\mathrm{hor}}_\pm$ are constants.  At spatial infinity,  i.e. when $r, r_* \to +\infty$ , the corresponding leading-order expressions are \cite{Langlois:2021xzq}
\begin{align}
	h_0(r) &= r (c^{\infty}_- e^{-i\omega r_*} - c^{\infty}_+ e^{i\omega r_*}) (1 + \mathcal{O}(1/r)) \,,\nonumber\\
	h_1(r) &= r (c^{\infty}_- e^{-i\omega r_*} + c^{\infty}_+ e^{i\omega r_*}) (1 + \mathcal{O}(1/r)) \,,
	\label{eq:asymp-schwa-infty}
\end{align}
where $c^{\infty}_\pm$ are other  constants and $r_* = r(1+ {o}(r))$ at infinity. 

Moreover, restoring the time dependence in $e^{-i\omega t}$, one sees that the asymptotic terms  in \eqref{eq:asymp-schwa-horiz} and \eqref{eq:asymp-schwa-infty}, consist of the superposition of an ingoing mode, proportional to $e^{-i\omega(t + r_*)}$, and of an outgoing mode, proportional to  $e^{-i\omega(t - r_*)}$,  propagating radially along the coordinate $r_*$ at speed 1. 

Quasinormal modes 
are the
perturbations that  are purely ingoing at the BH horizon and outgoing at spatial infinity, which is possible only for a discrete set of specific frequencies. Therefore, identifying  the QNMs means finding the complex values  $\omega$ such that the solution to \eqref{eq:first-order-schwarzschild-axial} 
satisfies
\begin{equation}
	c^{\infty}_- = 0 \qq{and} c^{\mathrm{hor}}_+ = 0 \,.
	\label{eq:BC-QNM-schwa}
\end{equation}
We will see in  the following subsection, how these QNM frequencies can be determined  numerically.

\subsection{Matrix continued fraction}
We show below how to 
adapt the continuous fraction method to compute QNMs directly from the first order system, without 
using
a Schr\"odinger-like reformation of the perturbations equations, {as in the seminal work by Leaver \cite{Leaver:1985ax}}.

\subsubsection{Boundary conditions and ansatz}

Imposing the QNM boundary conditions (\ref{eq:BC-QNM-schwa}) in the general asymptotic behaviours of $X_\mathrm{ax}$ and using $e^{i\omega r_*}=e^{i\omega r} (r-\mu)^{i\mu \omega}$, {which follows from (\ref{tortoise}),  we get}
\begin{eqnarray}
	X_\mathrm{ax}(r) & = &  c_-^{\mathrm{hor}} \, e^{-i\omega r} (r-\mu)^{-i\mu \omega}\begin{pmatrix} r-\mu  \\ \mu \end{pmatrix}  \left(\frac{1}{r-\mu}+  \mathcal{O}(1)\right)\,, \quad \text{when} \quad r \to \mu \, , 
	\label{eq:behav-schwa-horiz} \\
	X_\mathrm{ax}(r) & = & c_+^{\infty}  \, e^{i\omega r} (r-\mu)^{i\mu \omega}\begin{pmatrix}-1 \\ 1\end{pmatrix} (r+ {\cal O}(1))\,, \quad \text{when} \quad r \to +\infty \,.
\end{eqnarray}
In order to find a solution that satisfies simultaneously these two behaviours at the boundaries, one starts with an  ansatz of the form
\begin{equation}
	X_\mathrm{ax}(r) = e^{i\omega r} r^{1 + i\mu\omega} \qty(\frac{r-\mu}{r})^{-i\mu\omega} \times \begin{pmatrix} f_0(u) \\ f_1(u)/u \end{pmatrix} \qq{with} u = \frac{r-\mu}{r} \,.
	\label{eq:ansatz-Y}
\end{equation}
The  two complex-valued functions  $f_0$ and $f_1$  are supposed to be bounded (with no singularities) for $u \in [0,1]$ and  should satisfy the boundary conditions
 \begin{eqnarray}
 f_0(0)=f_1(0) \, , \qquad f_0(1)=-f_1(1) \, ,
 \end{eqnarray}
so that the appropriate asymptotic behaviours are indeed obtained. 

\subsubsection{Recurrence  relation}
\label{sec:gauss-elim-Schwa}

As the two functions are bounded for $u\in [0,1]$, they can be expanded  in power series as follows,
\begin{equation}
	f_0(u) = \sum_{n = 0}^{\infty} a_n u^n \qq{and} f_1(u) = \sum_{n = 0}^{\infty} b_n u^n \,,
	\label{eq:ansatz-f}
\end{equation}
where  $a_n$ and $b_n$ are complex numbers. Substituting the ansatz ~\eqref{eq:ansatz-Y}  together with the expressions~\eqref{eq:ansatz-f}   into the equations
of motion ~\eqref{eq:first-order-schwarzschild-axial} leads to a recurrence relation for the coefficients $a_n$ and $b_n$. In order to write this recurrence  relation in a compact form, it is
convenient to view the coefficients $a_n$ and $b_n$  as the components of  2-dimensional vectors $Y_n$, i.e.
\begin{eqnarray}
\label{defofYn}
Y_n = \begin{pmatrix} a_n \\ b_n \end{pmatrix} \,,
\end{eqnarray}
which in turn satisfy the relations
\begin{align}
	&\alpha_n Y_{n+1} + \beta_n Y_n + \gamma_n Y_{n-1} + \delta_n Y_{n-2} = 0 \, ,\qquad \forall n \geq 2 \,, 
	%& \alpha_1 Y_{2} + \beta_1 Y_1 + \gamma_1 Y_{0}  = 0 \,,\nonumber\\
	%& \alpha_0 Y_{1} + \beta_0 Y_0 = 0 \,,\nonumber\\
	\label{eq:rec-rel-schwa}
\end{align}
where the matrix coefficients are given  by
\begin{align}
	\alpha_n &= \begin{pmatrix} \frac{n+1 - i\mu\omega}{\mu} & i\omega\\ i\mu^2\omega  & \mu(n+1-i\mu\omega) \end{pmatrix} \,, &\beta_n =& \begin{pmatrix}\frac{-2n-1 + 4i\mu\omega}{\mu} & \flatfrac{-2i\lambda}{\mu^2\omega} \\ 0 & \mu(-2n+1+4i\mu\omega) \end{pmatrix} \,, \nonumber\\
	\gamma_n &= \begin{pmatrix} \frac{n}{\mu} - 2i\omega& \flatfrac{4i\lambda}{\mu^2\omega}\\ 0&\mu(n-2-2i\mu\omega) \end{pmatrix} \,, &\delta_n =& \begin{pmatrix}0 & \flatfrac{-2i\lambda}{\mu^2\omega} \\ 0 & 0 \end{pmatrix} \,. 
	\label{eq:coeffs-schwa-rec-rel}
\end{align}
This relation \eqref{eq:rec-rel-schwa} still holds for $0 \leq n < 2$ in which cases the number of terms is reduced,  
defining
$\alpha_n=\beta_n=\gamma_n=\delta_n=0$ when $n<0$ by convention.
When $n=0$,  it reduces to a 2-term  relation as it involves $Y_1$ and $Y_0$ only; when $n=1$, it is a 3-term relation 
between $Y_2$, $Y_1$ and $Y_0$.%

\medskip

In order to solve this 4-term recurrence  relation~\eqref{eq:rec-rel-schwa}, we first  show that  it  can always be reformulated
as a 3-term recurrence  relation.
To prove this is indeed possible, we proceed by induction. Hence, let us assume that it is possible to write~\eqref{eq:rec-rel-schwa} at some order $n$ in the form
\begin{equation}
	\tilde{\alpha}_n Y_{n+1} + \tilde{\beta}_n Y_n + \tilde{\gamma}_n Y_{n-1} = 0 \,,
	\label{eq:gauss-red-induction}
\end{equation}
where $\tilde\alpha_n$, $\tilde\beta_n$ and $\tilde\gamma_n$ are matrices to be determined, a priori different from \eqref{eq:coeffs-schwa-rec-rel}. If we assume, in addition,  that the matrix 
$\tilde\gamma_n$ is invertible, then we can express $Y_{n-1}$  as an explicit linear combination  of $Y_n$ and $Y_{n+1}$. As a consequence,  
the original 4-terms recurrence  relation~\eqref{eq:rec-rel-schwa} at order $n+1$ can also be reformulated as a 3-term recurrence  relation as follows,
\begin{equation}
	\alpha_{n+1} Y_{n+2} + \qty(\beta_{n+1} - \delta_{n+1} \cdot \tilde{\gamma}_n^{-1} \cdot \tilde{\alpha}_n) Y_{n+1} + \qty(\gamma_{n+1} - \delta_{n+1} \cdot \tilde{\gamma}_n^{-1} \cdot \tilde{\beta}_n) Y_{n}  = 0 \,.
\end{equation}
As~\eqref{eq:rec-rel-schwa} reduces to a 3-terms relation for $n=1$ (and also a 2-terms relation for $n=0$), it is indeed possible to transform the recurrence  relation into 
\eqref{eq:gauss-red-induction} for any $n$. The matrices entering in the 3-term recurrence  relation are recursively defined by
\begin{align}
	\tilde{\alpha}_n = \alpha_n \,,\quad
	\tilde{\beta}_n = \beta_{n} - \delta_{n} \cdot \tilde{\gamma}_{n-1}^{-1} \cdot \tilde{\alpha}_{n-1} \,,\quad
	\tilde{\gamma}_n = \gamma_{n} - \delta_{n} \cdot \tilde{\gamma}_{n-1}^{-1} \cdot \tilde{\beta}_{n-1} \,, \quad \text{for} \quad n \geq 1 \, ,
\end{align}
 with,  at order $n=0$, the initial matrices
 \begin{eqnarray}
 \tilde{\alpha}_0 = \alpha_0 \, , \qquad \tilde{\beta}_0 = \beta_0 \, , \qquad \tilde{\gamma}_0 = 0 \, .
 \end{eqnarray}

\subsubsection{Convergence of the series}

In general, three-term recurrence relations like~\eqref{eq:gauss-red-induction}  have solutions that can be expressed as a linear combination of two independent sequences (similarly to second order differential equations). However, not every solution will lead to the convergence of the power series~\eqref{eq:ansatz-f} whereas the 
convergence of 
both power series 
is required in order to impose the right boundary condition at infinity. In the case of scalar recurrence relations (where $Y_n$ are complex numbers), a condition  to ensure the convergence was found by Gautschi in~\cite{gautschiComputationalAspectsThreeTerm1967}, in the form of an equation containing a continued fraction. It was then used by Leaver~\cite{Leaver:1985ax} to compute QNMs, leading to the \enquote{continued fraction method}. 

Later, this study was generalised to recurrence relations  where $Y_n$ are vectors linked by matrix-valued coefficients  (see~\cite{simmendingerAnalyticalApproachFloquet1999, Rosa:2011my, Pani:2013pma} for instance). Here, we apply these results to our problem to ensure that
$f_0$ and $f_1$ are regular functions of $u$  in the whole interval $[0,1]$  with finite limits at the boundaries $u=0$ and $u=1$. This corresponds to choosing the appropriate branch for  $Y_n$. 

Let us proceed by first introducing  {\it invertible} matrices  $R_n$ such that
\begin{equation}
	Y_{n+1} = R_n Y_n \,, \qquad  \, n \geq 0 \, .
\end{equation}
Note that such relations do not uniquely define the matrices $R_n$. If one substitutes the 
above
relation into the recurrence relation \eqref{eq:gauss-red-induction}, one obtains
\begin{eqnarray}
(\tilde{\alpha}_0 R_0 + \tilde{\beta}_0) Y_0 = 0 \, , \qquad (\tilde \alpha_n R_n + \tilde \beta_n + \tilde \gamma_n R_{n-1}^{-1}) \, Y_n \; = \; 0  \qquad \text{for} \;\; n \geq 1 \,.
\label{eq:contfrac-order-0}
\end{eqnarray}
This identity is trivially satisfied if  the matrices $R_n$ themselves  satisfy  a recurrence relation which can be expressed as follows,
\begin{equation}
	R_{n-1} = -(\tilde{\beta}_n + \tilde{\alpha}_n R_n)^{-1} \tilde{\gamma}_n \, \qquad \text{for} \;\; n \geq 1 \, .
	\label{eq:contfrac-matrix}
\end{equation}
Interestingly,  when the matrices $R_n$ satisfy  this relation,  the power series defining the two functions $f_0$ and $f_1$ can be expected to be convergent.  
In other words, \eqref{eq:contfrac-matrix} selects the right branch for the solution $Y_n$. Furthermore, the zeroth order equation in \eqref{eq:contfrac-order-0}
can be seen as a system of algebraic equations for $\omega$ whose solutions are the quasi-normal frequencies of the Schwarzschild black hole.

Remarkably, one can construct  an infinite number of algebraic equations satisfied by the QNMs. Indeed, if we combine the $n=0$ order equation in \eqref{eq:contfrac-order-0}
with \eqref{eq:contfrac-matrix} for $n=1$, we obtain a new equation which links $R_1$ to $Y_1$,
\begin{eqnarray}
&& (\tilde{\alpha}_0 R_0 + \tilde{\beta}_0) Y_0 
	= (\tilde{\alpha}_0 R_0 + \tilde{\beta}_0) R_0^{-1} R_0 Y_0 = \qty[\tilde{\alpha}_0 - \tilde{\beta}_0 \tilde{\gamma}_1^{-1} (\tilde{\beta}_1 + \tilde{\alpha}_1 R_1) ]Y_1 =0 \nonumber \\
  &&	\qquad  \qquad \Longrightarrow \qty[\tilde{\alpha}_1 R_1 + \tilde{\beta}_1 - \tilde{\gamma}_1 \tilde{\beta}_0^{-1} \tilde{\alpha}_0 ]Y_1 = 0 \,.
\end{eqnarray}
It can be viewed again as an algebraic equation for the QNMs. Of course, we can proceed in this way recursively
 to obtain, at any order $n \geq 0$,  
 relations of the form
 \begin{align}
	(\tilde{\alpha}_n R_n + Q_n) Y_n = 0 \,,\label{eq:contfrac}
\end{align}
 where the matrices $Q_n$ must satisfy
 \begin{align}	
	\text{for} \, n \geq 1 ,\quad Q_n = \tilde{\beta}_n - \tilde{\gamma}_n \cdot Q_{n-1}^{-1}\cdot \tilde{\alpha}_{n-1} \qquad \text{and} \quad 
	Q_0 = \tilde{\beta}_0 \,.
	\label{eq:contfrac-Qn}
\end{align}
Hence, the method  is reminiscent of the usual construction of a continued fraction based on the Schr\"odinger-like formulation. 
At each order, the equation~\eqref{eq:contfrac} can  be seen as an algebraic equation for $\omega$ and then, each solution of this equation 
corresponds to a QNM. 
Of course, there exists a non-trivial solution of eq.~\eqref{eq:contfrac} only when 
\begin{equation}
	\det(\tilde{\alpha}_n R_n + Q_n) = 0 \,, 
	\label{eq:contfrac-det}
\end{equation} 
which is the equation we will solve numerically in the following section. {In principle, \eqref{eq:contfrac-det} is necessary but not sufficient to satisfy  \eqref{eq:contfrac}. Therefore, we will need to check that the values of $\omega$ we obtain by solving the former equation also solve the latter\footnote{This is done in section \ref{sec:check-Y0}.}. Further details on the numerical resolution will be given in the next subsection \ref{sec:QNM-Schwa}.

\subsection{Numerical method and results}
\label{sec:QNM-Schwa}

The numerical method consists in applying a root-finding algorithm in the complex plane to determine the values of $\omega$ that solve \eqref{eq:contfrac-det}, for given values of the black hole mass $2\mu$, of the angular momentum integer $\ell$, or equivalently $\lambda$, and the so-called inversion index $n$.

One must first determine the matrix  $R_n$, which can be computed from  $R_{n+1}$ via the inverse recurrence relation (\ref{eq:contfrac-matrix}). In practice,  in order  to compute $R_n$ in a finite number of steps, a truncation is necessary at some large value $N$, where we impose the simple condition
\begin{equation}
\label{truncation}
	R_N = 0 \,.
\end{equation}
The precision on the quasi-normal frequency  is then controlled by the value of the truncation index $N$. 
One should note that the value we choose for $R_N$ is arbitrary: it will introduce a small error which 
turns out to be negligible when $N$ is chosen to be large enough.

The other matrix appearing in \eqref{eq:contfrac-det}, $Q_n$, is computed via the recurrence relation \eqref{eq:contfrac-Qn}. The last step consists in finding  the frequencies solving \eqref{eq:contfrac-det}  thanks to  the root-finding algorithm. One finally checks that the values thus obtained are stable when the truncation integer $N$ is increased. 

We find empirically that increasing $n$ leads to a better precision for the computation of high-overtone modes. By choosing increasingly large values for $N$, we are able to compute several hundreds of QNMs for the Schwarzschild black hole, up to very high precision (around 10 digits). More details on the convergence and consistency checks are given in section~\ref{sec:consistency} in the case of the BCL black hole. Since the Schwarzschild QNMs are well-known, we simply provide  the first 20 frequencies obtained by our method in table~\ref{tab:Schwarzschild-QNMs}. These values agree with existing data (for example~\cite{Berti:2009kk}) up to the  6-digit precision we set for our computation, which confirms the validity of  our method.

{
\renewcommand{\arraystretch}{1.15}
\begin{table}[!htb]
	\begin{minipage}{0.5\columnwidth}
		\begin{tabular}{|>{\centering\arraybackslash}p{1cm}|>{\centering\arraybackslash}p{4cm}|}
			\hline
			$\bm{n}$ & $\bm{\omega_n}$ \\\hline\hline
			0 & 0.747343-0.177924i \\
			1 & 0.693422-0.547830i \\
			2 & 0.602107-0.956554i \\
			3 & 0.503010-1.410296i \\
			4 & 0.415029-1.893690i \\
			5 & 0.338598-2.391216i \\
			6 & 0.266504-2.895821i \\
			7 & 0.185645-3.407682i \\
			8 & 0.000000-3.999000i \\
			9 & 0.126527-4.605289i \\
			\hline
		\end{tabular}
	\end{minipage}%
	\begin{minipage}{0.5\columnwidth}
		\begin{tabular}{|>{\centering\arraybackslash}p{1cm}|>{\centering\arraybackslash}p{4cm}|}
			\hline
			$\bm{n}$ & $\bm{\omega_n}$ \\\hline\hline
			10 & 0.153107-5.121653i \\
			11 & 0.165196-5.630884i \\
			12 & 0.171456-6.137389i \\
			13 & 0.174788-6.642460i \\
			14 & 0.176478-7.146641i \\
			15 & 0.177181-7.650211i \\
			16 & 0.177265-8.153329i \\
			17 & 0.176953-8.656100i \\
			18 & 0.176381-9.158594i \\
			19 & 0.175641-9.660860i \\
			\hline
		\end{tabular}
	\end{minipage}
	\caption{\small{Values computed for the QNMs of the Schwarzschild black hole using the first order system and the matrix continued fraction method. We set $\mu = 1$ and $\lambda = 2$ (corresponding to $\ell = 2$). We focus only on modes with $\Re(\omega) > 0$; for each mode in this half-plane, there exists a mode with the same imaginary part and an opposite real part.}}
	\label{tab:Schwarzschild-QNMs}
\end{table}
}

\section{BCL black hole in Scalar-Tensor Theories}
We now consider an exact  black hole solution in a scalar-tensor theory of gravity, for which we compute for the first time the axial quasi-normal modes,
as presented in the next section.

\subsection{BH solution in modified gravity}

\label{sec:BCL}
The scalar-tensor theory discussed here is 
described by an action of the form 
\begin{equation}
	S[g_{\mu\nu},\phi] = \int \dd^4{x} \sqrt{-g} \Big[F(X) R + P(X) %+ Q(X,\phi) \square\phi 
	+ 2\pdv{F}{X}(X, \phi) (\phi_{\mu\nu}\phi^{\mu\nu} - (\square\phi)^2) \Big]\, ,
	\label{eq:action-Horndeski}
\end{equation}
where $X\equiv \nabla^\mu\phi\nabla_\mu\phi$, $\phi_{\mu\nu}\equiv \nabla_\mu\nabla_\nu\phi$ and
\begin{equation}
	F(X) = F_0 + F_1 \sqrt{X} \,,\qquad P(X) = - P_1 X  \,.
	\label{eq:choice-Horndeski-BCL}
\end{equation}
This action, which depends on second derivatives of the scalar field $\phi$, belongs to the  subfamily of Horndeski theories~\cite{Horndeski:1974wa}, itself included in the general family DHOST theories discussed in the introduction.

Our motivation for choosing this specific theory is the existence of an exact BH solution, 
dubbed BCL solution after its authors \cite{Babichev:2017guv}. It is described by the metric 
\begin{equation}
	\dd{s}^2 = - A(r) \dd{t}^2 + \frac{1}{B(r)} \dd{r}^2 + C(r) \dd{\Omega}^2\,,
\end{equation}
with
\begin{equation}
A(r) = B(r) = \qty(1 - \frac{r_+}{r}) \qty(1 + \frac{r_-}{r}) \,,\qquad C(r) = r^2\,,
\label{eq:A-B-C-BCL}	
\end{equation}
where the (positive) quantities $r_+$ and $r_-$ are defined by
\begin{equation}
\label{rpm}
	r_+ r_- = \frac{F_1^2}{2F_0P_1} \,,\quad r_+ - r_- = \mu \qq{and} r_+ > r_- > 0 \,.
\end{equation}
 The metric possesses only one horizon located at $r = r_+$  and the function $A$ can be rewritten as
\begin{equation}
	A(r) = 1 - \frac{\mu}{r} - \frac{\mu^2 \xi}{2r^2} \qq{with} \xi = \frac{2 r_+ r_-}{\mu^2} \,,
\end{equation}
which shows that   $\mu$ corresponds to twice the ADM mass, as previously in the Schwarzschild case. 

As for the scalar field, its  configuration is given by
\begin{equation}
\phi(r)=\pm \frac{\bb}{\gp\sqrt{\rp\rz}}\arctan\left[\frac{(r_+ - r_-)r + 2\rp\rz}{2\sqrt{ \rp\rz}\sqrt{(r-\rp)(r+\rz)}}\right]\, + {\text{cst}} \, .
\end{equation}
The global sign of $\phi(r)$ and the constant are physically irrelevant \cite{Babichev:2017guv}. 

Interestingly, the above solution can be seen as a one-parameter deformation of the Schwarzschild solution, with $r_-$ playing the role of the 
deformation
parameter. In the limit $r_-=0$, corresponding to $F_1=0$, one recovers precisely the Schwarzschild metric, while the scalar field vanishes.

\subsection{Perturbation equations}
To obtain the equations of motion for the linear perturbations, one proceeds similarly to the GR case presented in  section~\ref{sec:GR-perturbations}. We substitute in the action \eqref{eq:action-Horndeski} the metric and scalar field 
\begin{align}
	g_{\mu\nu} &= \bar{g}_{\mu\nu} + h_{\mu\nu} \,,\qquad
	\phi = \bar{\phi} + \var{\phi} \,,
\end{align}
where $\bar{g}_{\mu\nu}$ and $\bar{\phi}$ are respectively the metric and scalar field of the background, while  $h_{\mu\nu}$  and $\var{\phi}$ are the corresponding perturbations.

 We 
 then expand
  the action~\eqref{eq:action-Horndeski} up to quadratic order in $h_{\mu\nu}$ and $\delta\phi$, and obtain (after some calculations) the quadratic action for  linear perturbations
\begin{equation}
	S_\mathrm{quad}[h_{\mu\nu}, \var{\phi}] \,.
\end{equation}
The equations of motion for the perturbations are then given by the Euler-Lagrange equations,
\begin{equation}
	\mathcal{E}_{\mu\nu} = \frac{\delta S_\mathrm{quad}}{\delta h_{\mu\nu}} = 0 \qq{and} \mathcal{E}_{\phi} = \frac{\delta S_\mathrm{quad}}{\delta (\var{\phi})} = 0 \,.
\end{equation}
One can check that the equation $\mathcal{E}_{\phi} = 0$ is redundant due to Bianchi's identities.  

In the following,  we only consider  axial perturbations, for which the perturbation of the scalar field vanishes, i.e. $\var{\phi}=0$. The situation is then analogous to the GR case, although the equations of motion are now different. 
As shown in \cite{Langlois:2021aji}, 
we obtain the first-order system
\begin{equation}
	\dv{X_\mathrm{ax}}{r} = M_\mathrm{ax} X_\mathrm{ax} %\qq{and} \dv{X_\mathrm{po}}{r} = M_\mathrm{po} X_\mathrm{po}  \,,
	\label{eq:first-order-systs}
\end{equation}
with 
\begin{equation}
	X_\mathrm{ax} = \begin{pmatrix} h_0 \\ h_1 \end{pmatrix}\,, \qquad \qquad M_\mathrm{ax} = \begin{pmatrix}
		\frac{2}{r} & -i\omega + \frac{2i\lambda(r-r_+)(r+r_-)}{r^4 \omega} \\
		-\frac{i\omega r^2(r^2+2r_-r_+)}{(r-r_+)^2 (r-r_-)^2} & - \frac{r(r_+ - r_-) + 2r_+r_-}{r(r-r_+)(r+r_-)} \, 
	\end{pmatrix} \, .
	\label{eq:first-order-BCL}
\end{equation}
As we can see, we recover the Schwarzschild system \eqref{eq:first-order-schwarzschild-axial} in the limit $r_-=0$, where $\mu=r_+$.

\section{Axial quasinormal modes of the BCL Black Hole}
\label{sec:QNMs-BCL}

In this section, we compute  the axial QNMs of the BCL black hole using  the \enquote{matrix continued fraction} method introduced earlier for Schwarzschild. Then, we compare our results with those obtained from  already existing methods to perform consistency checks.

\subsection{Ansatz and recurrence relation}
In order to guess an appropriate ansatz,
 we need the asymptotic behaviours of axial perturbations at both the horizon and spatial infinity. They have already been computed  in~\cite{Langlois:2021aji} where we found that
 the leading-order  terms in the asymptotic expansion  at spatial infinity (when $r \to \infty$) are given by
\begin{align}
	h_0(r) &=r (c^{\infty}_+ e^{i\omega r}r^{i\mu\omega} - c^{\infty}_- e^{-i\omega r}r^{-i\mu\omega}) (1+ \mathcal{O}(1/r)) \,,\nonumber\\
	h_1(r) &= r(c^{\infty}_+ e^{i\omega r}r^{i\mu\omega} + c^{\infty}_- e^{-i\omega r}r^{-i\mu\omega}) (1 + \mathcal{O}(1/r)) \,,
	\label{eq:asymp-BCL-infty}
\end{align}
where $c^{\infty}_\pm$ are constant. And the asymptotic expansion near  the horizon (when $r \to r_+$) yields, at leading order,
\begin{align}
	h_0(r) &= (c^{\mathrm{hor}}_+ (r-r_+)^{+i\omega r_0} + c^{\mathrm{hor}}_- (r-r_+)^{-i\omega r_0}) (1 + \mathcal{O}(r-r_+)) \,,\nonumber\\
	h_1(r) &= \frac{r_0}{r_+} (-c^{\mathrm{hor}}_+ (r-r_+)^{+i\omega r_0-1} + c^{\mathrm{hor}}_- (r-r_+)^{-i\omega r_0-1}) (1 + \mathcal{O}(r-r_+)) \,,
	\label{eq:asymp-BCL-horiz}
\end{align}
where $c^{\mathrm{hor}}_\pm$ are constant and $r_0$, which has the dimension of a radius, is defined by 
\begin{equation}
	r_0 = r_+ \frac{\sqrt{r_+ (r_+ + 2 r_-)}}{r_+ + r_-} \,.
	\label{eq:def-r0}
\end{equation}

As usual, QNMs are outgoing at infinity and ingoing at the horizon, corresponding to the boundary conditions  $c^{\infty}_- = 0$ and $c^{\mathrm{hor}}_+ = 0$. Following the same strategy as in the GR case, we choose the following ansatz for the solution,%
\begin{equation}
	X_\mathrm{ax}(r) = e^{i \omega r} r^{i \mu \omega + 1} \qty(\frac{r-r_+}{r})^{-i \omega r_0} \begin{pmatrix} f_0(u) \\ f_1(u)/u \end{pmatrix}  \qq{with} u = \frac{r-r_+}{r} \,,
	\label{eq:ansatz-BCL-axial}
\end{equation}
where $f_0$ and $f_1$ are supposed to be bounded in the whole domain $u \in [0,1]$ and should satisfy the boundary conditions
\begin{equation}
	f_0(0) = \frac{r_+}{r_0} f_1(0) \qq{and} f_0(1) = - f_1(1) \,,
	\label{eq:BC-fi-BCL}
\end{equation}
We again decompose the two functions $f_0$ and $f_1$  in power series,
\begin{equation}
	f_0(u) = \sum_{n = 0}^{\infty} a_n u^n \qq{and} f_1(u) = \sum_{n = 0}^{\infty} b_n u^n \,,
	\label{eq:power-series-BCL}
\end{equation}
and the differential system \eqref{eq:first-order-systs} leads to a recurrence relation for the vector $Y_n$, defined as  in \eqref{defofYn}, that now involves 5 terms,
\begin{align}
		&\alpha_n Y_{n+1} + \beta_n Y_n + \gamma_n Y_{n-1} + \delta_n Y_{n-2} + \varepsilon_n Y_{n-3}= 0 \,, \qquad n \geq 3 \, .
	\label{eq:rec-rel-BCL}
\end{align}
The matrix coefficients are given explicitly in  appendix \ref{app:rec-rel-BCL}. This relation still holds for $0 \leq n < 3$ in which cases the number of terms is reduced, defining $\alpha_n=\beta_n=\gamma_n=\delta_n=\epsilon_n=0$ when $n<0$ by convention.
When $n=0$,  it reduces to a 2-terms  relation as it involves $Y_1$ and $Y_0$ only; when $n=1$, it is a 3-terms relation 
(between $Y_2$, $Y_1$ and $Y_0$);  when $n=2$, it is a 4-terms relation  (between $Y_3$, $Y_2$, $Y_1$ and $Y_0$). 

The process of casting the recurrence  relation~\eqref{eq:rec-rel-BCL} into a three-term recurrence  relation is completely similar to the one presented in section \ref{sec:gauss-elim-Schwa}, with an additional  step since the relation in the BCL case is five-term long. It is therefore always possible to recover a recurrence  relation of the form~\eqref{eq:gauss-red-induction}, with values of the matrix coefficients $\tilde{\alpha}_n$, $\tilde{\beta}_n$ and $\tilde{\gamma}_n$ depending on $r_+$ and $r_-$. The expressions of these matrices  are given in  appendix \ref{app:gauss-reduc-BCL}.

\subsection{Numerical results}
\label{sec:numerical-results}

In this subsection, we present the QNM frequencies obtained  by our numerical method  applied to the BCL axial perturbations. Convergence tests and consistency checks (via comparison with other references or methods) will be discussed in section \ref{sec:consistency} { and in appendix \ref{app:other-methods}}.

As explained above, the QNM spectrum is obtained from a resolution of  \eqref{eq:contfrac-det} with the truncation \eqref{truncation} at a given rank $N$.
We use a root-finding method  to locate the position of the modes and the threshold for convergence is set at variations smaller than $10^{-6}$, unless stated otherwise.

We work in units of $r_+$, which corresponds to setting $r_+ = 1$. On figure~\ref{fig:qnm-spectrum-axial} where $\lambda = 2$ (which corresponds to $\ell = 2$), we show the first 40 modes and their 
migration in the complex plane
when $r_-$ goes from 0, which corresponds to the Schwarzschild case, to $0.5$.  Hence, 
as already mentioned, $r_-$ parametrises the deviation from GR. We have 
also noticed that the evolution of these modes with $r_-$ 
is very similar for higher values of lambda.
For this  reason we do not show the QNM spectra for other values of $\lambda$. 
\begin{figure}[!htb]
	\centering
	\includegraphics{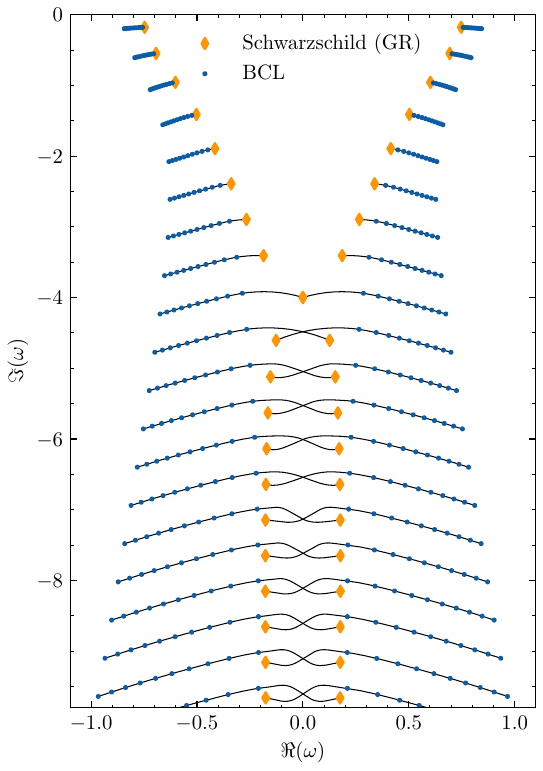}
	\caption{\small Axial quasinormal mode spectrum of the BCL black hole, for $\lambda = 2$ and $r_+ = 1$. The parameter $r_-$ varies between 0 (Schwarzschild case, represented by diamonds) and 0.5. Blue dots correspond to increments of $0.05$ to the value of $r_-$.}
	\label{fig:qnm-spectrum-axial}
\end{figure}

We observe that higher-overtone modes are much more sensitive to small variations of $r_-$ (and thus to deviations from GR) even though
 this does not seem to come from a spectral  instability   \cite{Jaramillo:2020tuu} here.
For high overtones, the QNM points appear to be aligned    along a straight line which, in contrast  with the Schwarzschild case, is no longer vertical when $r_-\neq 0$. We can check that modes of much higher overtone still follow the same line, which seems to indicate  that it corresponds to a real asymptote of the QNM spectrum. 
We have not been able to find an analytical argument to explain this behaviour.
On figure~\ref{fig:qnm-asymptote-axial}, we illustrate the existence of such an asymptote at high overtones, which can be parametrised by the equation
\begin{equation}
	\Im(\omega) = a \times \Re(\omega) + b \,.
\end{equation}
\begin{figure}[!htb]
	\centering
	\includegraphics{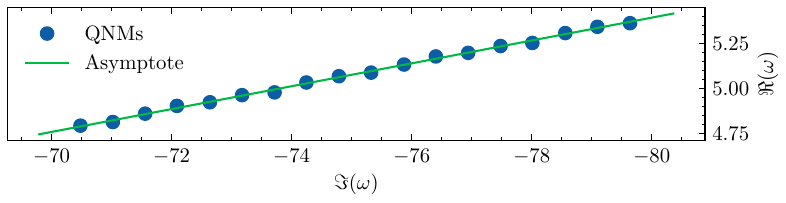}
	\caption{\small Asymptote of the QNM spectrum for $\lambda = 2$, $r_+ = 1$ and $r_- = 0.5$. The real and imaginary axes are rotated by 90 degrees compared to figure~\ref{fig:qnm-spectrum-axial} for the sake of clarity.}
	\label{fig:qnm-asymptote-axial}
\end{figure}
The dependence on $r_-$ of
the slope of the asymptote 
is shown in figure~\ref{fig:slope-asymptote-axial}. 
The Schwarzschild case $r_- = 0$ corresponds to a vertical asymptote with $1/a = 0$. 
As $r_-$ becomes larger, the asymptote  is less and less vertical.

\begin{figure}[!htb]
	\centering
	\includegraphics{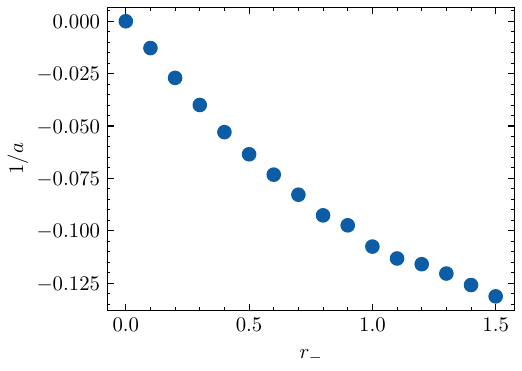}
	\caption{\small Inverse of the slope $a$ of the asymptote for $r_+ = 1$ and $\lambda = 2$. Values of $r_-$ span between 0 (Schwarzschild case where the asymptote is vertical and $1/a = 0$) and 1.5, with increments of 0.1.}
	\label{fig:slope-asymptote-axial}
\end{figure}

As a final remark, we recall that the Schwarzschild QNM spectrum is well-known to contain special  modes which have a vanishing real part, 
meaning
they are purely-damped (non-oscillating) modes.  They are associated with an \enquote{algebraically special} solution of the perturbation equations, obtained in \cite{chandrasekharAlgebraicallySpecialPerturbations1997}, and linked to the Robinson-Trautman metric describing nonperturbative gravitational waves onto a Schwarzschild background \cite{Qi:1993ey}. When $\ell=2$,  there is one such mode which corresponds to the overtone $n = 8$ on figure~\ref{fig:qnm-spectrum-axial}.
In the case of the BCL black hole, one sees that the mode of overtone $n = 8$ is no longer purely-damped: as $r_-$ becomes non-zero, this mode splits into two modes of positive and negative real parts. Nevertheless, one can see that every subsequent {mode} moves towards  the imaginary axis  when $r_-$ increases and, for some {critical} value of $r_-$, reaches this axis. As a consequence, for some particular values of $r_-$, algebraically special modes still exist for the BCL black hole.

\section{Consistency checks}
\label{sec:consistency}

The numerical resolution of \eqref{eq:contfrac-det}    requires a truncation of the recurrence  relations at some finite rank $N$ \eqref{truncation}. Such a truncation
enables us to compute the matrices $R_n$ for any $n \leq N$ and then to solve \eqref{eq:contfrac-det}.

\subsection{Convergence of the method when $N$ increases}
As a first consistency check, we need to make sure that the QNM frequencies thus  computed converge towards a fixed value when $N$ increases. This is shown on figure~\ref{fig:convergence} where we plot the difference between the computed QNM value $\omega$ in the Schwarzschild case ($r_- = 0$) of axial perturbations and the known value $\omega_*$ computed in~\cite{Berti:2009kk}. We study several values of the overtone $n$. We observe that as $N$ increases, all modes converge to their theoretical value; however, higher overtones need higher values of $N$ 
to reach a good precision.

\begin{figure}[!htb]
	\centering
	\includegraphics{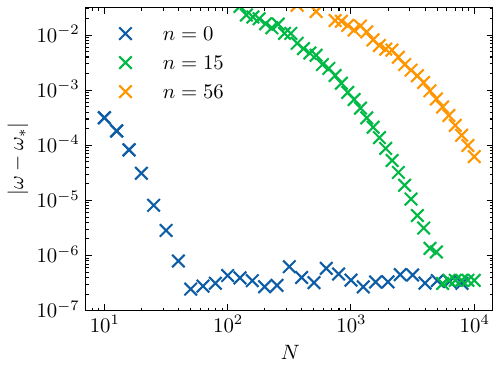}
	\caption{\small Convergence of the computed QNM frequencies, for different overtones. The convergence threshold is set at $10^{-6}$, which explains why the convergence does not improve beyond this value.}
	\label{fig:convergence}
\end{figure}

\subsection{QNM mode functions}

Given a value $\omega$ in the QNM spectrum, it is possible to perform a self-consistency check by computing $f_0$ and $f_1$ in \eqref{eq:power-series-BCL} and verifying that the resulting functions solve the perturbation equations with the required boundary conditions as expected.

To see this is indeed the case, we first reformulate the perturbation equations in terms of $f_0$ and $f_1$.
Hence, we substitute the ansatz \eqref{eq:ansatz-BCL-axial} into the original perturbation equations (\ref{eq:first-order-systs}-\ref{eq:first-order-BCL}), and we obtain that $f_0$ and $f_1$ must satisfy
\begin{eqnarray}
        \mathcal{E}_0 &\equiv& g_0'(r) + \kappa_1(r) g_0(r) + \kappa_2(r) g_1(r) = 0\,,\nonumber\\
	\mathcal{E}_1 &\equiv& g_1'(r) + \kappa_3(r) g_0(r) + \kappa_4(r) g_1(r)  =0 \,,
	\label{eq:eqs-axial-ansatz}
\end{eqnarray}
where $g_i(r) = f_i(u(r))$ and the functions $\kappa_i$ are given by
\begin{align}
	\kappa_1(r) &= \frac{1}{r} \qty[-1 + i \omega \qty(r - r_- + r_+ - \frac{r_+^2 r_0}{r-r_+})]\,, \nonumber \\
	\kappa_2(r) &= \frac{i r \omega}{r - r_+} - 2 i \lambda \frac{r + r_-}{r^3 \omega}\,,\nonumber\\
	\kappa_3(r) &= \frac{i r (r^2 + 2 r_+ r_-)}{(r-r_+) (r+r_-)^2}\,, \nonumber \\
	\kappa_4(r) &= \frac{1}{r+r_-} + i\omega \qty[1 + \frac{1}{r}\qty(r_+ - r_- - \frac{r_+^2 r_0}{r - r_+})]\,.
\end{align}
We recall that the constants $r_\pm$ and $r_0$  are defined in \eqref{rpm} and  in \eqref{eq:def-r0}, respectively.
Once we have these equations, we wish to check that, when $\omega$ belongs to the QNM spectrum, 
the functions $f_0$ and $f_1$ obtained numerically are indeed solutions. We perform two tests: first we pick up a frequency which is not a QNM and then we compare to the case where $\omega$ belongs to the spectrum.

 \medskip
 
Let us start by considering an arbitrary value of $\omega$. We compute the coefficients $a_n$ and $b_n$ and 
truncate the power series in~\eqref{eq:ansatz-f} at some order $M$. We plot in figure~\ref{fig:plot-axial-example} the values
of $a_n$ and $b_n$ and see that they do not tend to zero, but on the contrary  seem to diverge, which implies that the series ~\eqref{eq:ansatz-f}  are ill-defined. 
Furthermore, whereas equations $\mathcal{E}_0 = 0$ and $\mathcal{E}_1 = 0$ are verified close to the horizon, which is consistent with the fact that the power series decomposition of $f$ is done with respect to the variable $u=\flatfrac{(r-r_+)}{r}$, while  at high values of $r$, the equations are clearly not satisfied very far from zero as $\mathcal{E}_0$ and $\mathcal{E}_1$ reach a constant value . As a consequence, when $\omega$ is not a QNM frequency, the numerical method does not lead to a solution for $f_0$ and $f_1$.

\begin{figure}[!htb]
	\centering
	\begin{subfigure}[t]{0.45\textwidth}
		\centering
		\vspace{0pt}
		\includegraphics{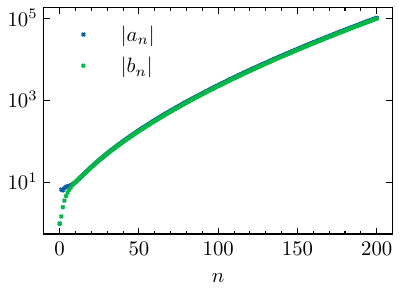}
		\label{fig:plot-coeffs-axial-example}
	\end{subfigure}
	\hspace{0.5cm}
	\begin{subfigure}[t]{0.45\textwidth}
		\centering
		\vspace{0pt}
		\includegraphics{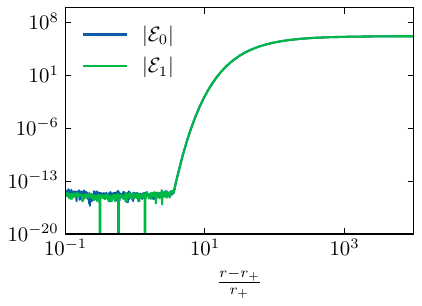}
		\label{fig:plot-eqs-axial-example}
	\end{subfigure}
	\caption{\small Absolute value of the coefficients $a_{n}$ and $b_{n}$ along with the absolute value of equations $\mathcal{E}_0$ and $\mathcal{E}_1$ for $r_+ = 1$, $r_- = 0.2$, $\lambda = 2$ and $\omega = 0.5 - 0.3i$. The truncation is taken at $M = 200$.}
	\label{fig:plot-axial-example}
\end{figure}

Let us now study the situation where $\omega$ belongs to the QNM spectrum. As an example, we take the fundamental mode $\omega_0$ for $\ell=2$ when $r_- = 0.2$. 
The numerical calculation gives
\begin{eqnarray}
	\omega_0 = 0.785460 - 0.184148i\,\quad \text{or} \quad \,
	\omega_0= 0.78546018859393 - 0.18414793488781i\,,
\end{eqnarray}
when the convergence threshold is put at $10^{-6}$ or  at $10^{-14}$ respectively. Then, we compute the coefficients $a_n$ and $b_n$ for these two different precisions. The
results are plotted in figure \ref{fig:plot-axial-coeffs-threshold}. We see that the first coefficients do seem to converge to zero, at least up to a certain value of $n$ represented by a dotted vertical line. As the accuracy of the mode increases, the convergence of the sequences $a_n$ and $b_n$ improves. Henceforth, we should truncate the sequence at the point where the coefficients start diverging: this means that with 6-digit accuracy, we should take $M \simeq 20$ while using 14-digit accuracy, we should use $M \simeq 100$.

\begin{figure}[!htb]
	\centering
	\begin{subfigure}[t]{0.45\textwidth}
		\centering
		\vspace{0pt}
		\includegraphics{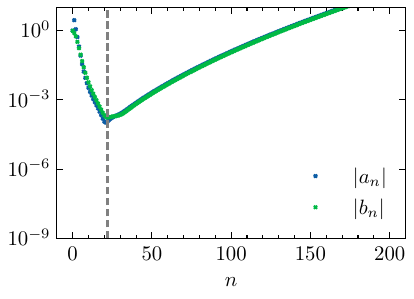}
		\caption{6-digits accuracy}
		\label{fig:plot-axial-coeffs-threshold-6}
	\end{subfigure}
	\hspace{0.5cm}
	\begin{subfigure}[t]{0.45\textwidth}
		\centering
		\vspace{0pt}
		\includegraphics{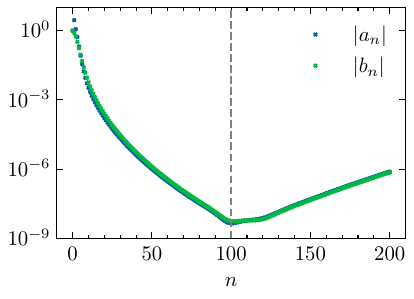}
		\caption{14-digits accuracy}
		\label{fig:plot-axial-coeffs-threshold-14}
	\end{subfigure}
	\caption{\small Absolute values of the coefficients $a_{n}$ and $b_{n}$ for $r_+ = 1$, $r_- = 0.2$, $\lambda = 2$ and $\omega = \omega_0$. We consider two cases: in~\ref{fig:plot-axial-coeffs-threshold-6}, we compute the value of $\omega_0$ up to 6 decimals and in~\ref{fig:plot-axial-coeffs-threshold-14}, we compute it up to 14 decimals. In both cases, the truncation is taken at $M = 200$. The vertical dotted line corresponds to threshold values of $n$ above which the coefficients do not seem to converge to zero anymore.}
	\label{fig:plot-axial-coeffs-threshold}
\end{figure}

Finally we evaluate   $\mathcal{E}_0$ and $\mathcal{E}_1$ to see whether the two series are indeed solutions of the equations. As in the previous case, these two equations
are satisfied near the horizon. It is more interesting to look at these equations in the limit $r \rightarrow \infty$ where both equations converge towards a constant value at infinity similarly to what was seen in figure~\ref{fig:plot-axial-example} for a non-QNM value of $\omega$. Furthermore, we notice that the asymptotic values for $\abs{\mathcal{E}_0}$ and $\abs{\mathcal{E}_1}$ are very close, and then we define $\mathcal{E}_\infty$ by
\begin{equation}
	\mathcal{E}_\infty = \lim_{r\to\infty} \abs{\mathcal{E}_0} \simeq \lim_{r\to\infty} \abs{\mathcal{E}_1} \,.
\end{equation}
On figure~\ref{fig:plot-axial-eqs-threshold}, we plot the values of $\mathcal{E}_\infty$ for different truncation values $M$. We find that, at low truncation rank $M$, $\mathcal{E}_\infty$ decreases with $M$ and reaches very low values. This means that the equations are solved and the value of $\omega$ is indeed a QNM. At higher values of $M$ however, equations do not seem to converge to zero anymore: this is because the sequences of coefficients $a_n$ and $b_n$ have stopped converging (as one can see in figure~\ref{fig:plot-axial-coeffs-threshold}). What is important is that this threshold on $M$ increases as the precision on the QNM gets better: an arbitrarily high precision on the QNM frequency would lead to fully converging series of $a_n$ and $b_n$, and therefore a monotonous behaviour of $\mathcal{E}_\infty$.

\begin{figure}[!htb]
	\centering
	\includegraphics{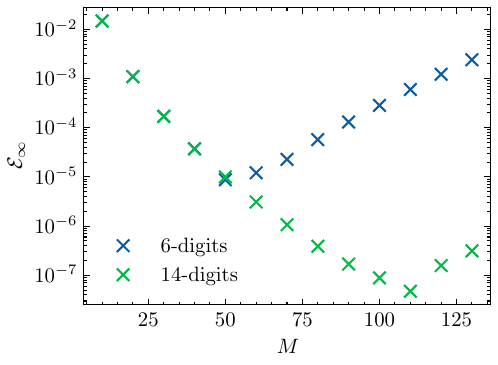}
	\caption{\small Values of $\mathcal{E}_\infty$ for $r_+ = 1$, $r_- = 0.2$, $\lambda = 2$ and $\omega = \omega_0$, for varying values of the truncation. We see that values decrease as $M$ increases, which means that the perturbation equations are verified. When $M$ becomes too high, convergence stops. We see that this threshold on $M$ increases with the precision sought for the NM value: as the precision increases, the threshold also increases.}
	\label{fig:plot-axial-eqs-threshold}
\end{figure}

\medskip

In conclusion, we have shown that the QNM frequency $\omega$ computed with our method is associated with mode functions that do obey the perturbation equations~\eqref{eq:eqs-axial-ansatz}. As the truncation value of the series~\eqref{eq:ansatz-f} increases, we have checked that equations $\mathcal{E}_0$ and $\mathcal{E}_1$ get closer to zero, at least up to a threshold determined by the accuracy with which the QNM frequency was computed. This constitutes a strong self-consistency check of our method.

\subsection{Initial coefficient vs. boundary condition at the horizon}
\label{sec:check-Y0}

Each QNM computed is such that the determinant of $\tilde{\alpha}_0 R_0 + Q_0$ is zero, which means that this matrix admits at least one null direction $Y_0$ \eqref{eq:contfrac-order-0}. However, $Y_0$ is not arbitrary and it has been fixed by the asymptotic behaviour of the differential system at the horizon, given by \eqref{eq:BC-fi-BCL} for the BCL black hole, which leads to
\begin{equation}
	Y_0 \propto  \begin{pmatrix}r_+ \\ r_0\end{pmatrix} \,.
\end{equation}
In order to check this property, we compute  a null vector $U$  of $\tilde{\alpha}_0 R_0 + Q_0$ for each QNM we have identified. To compare the directions of $U$ and $Y_0$, we  compute their determinant or, equivalently, the quantity $\eta$  defined by
\begin{equation}
	\eta = \frac{U_1}{U_0} - \frac{r_0}{r_+} \,.
\end{equation}
The results are given in  table \eqref{tab:check-YO} for different QNMs and different BH parameters. One can observe that the numerical  vector $U$ is colinear with theoretical null vector $Y_0$, which is another strong indication of the consistency of our method. Notice however that the agreement worsens as the overtone increases. Interestingly, it has been observed (see \cite{Leaver:1985ax}) that solving the equation
\begin{equation}
	\det(\tilde{\alpha}_m R_m + Q_m) = 0
	\label{eq:eq-det-BCL-inverted}
\end{equation}
for $m>0$ gives a better precision for the computation of the overtone $m$. Then, in order to have a good precision for high overtones, one should instead solve \eqref{eq:eq-det-BCL-inverted}  for the corresponding value of $m$.

{
	\renewcommand{\arraystretch}{1.15}
	\begin{table}[!htb]
		\begin{minipage}{0.3\columnwidth}
			\begin{tabular}{|>{\centering\arraybackslash}p{1cm}|>{\centering\arraybackslash}p{2cm}|}
				\hline
				$\bm{n}$ & $\bm{\eta}$ \\\hline\hline
				0 &  \num{1.1e-6}\\
				1 &  \num{1.5e-6}\\
				2 &  \num{4.3e-6}\\
				3 &  \num{3.8e-6}\\
				4 &  \num{1.2e-5}\\
				5 &  \num{3.4e-5}\\
				6 &  \num{6.2e-5}\\
				7 &  \num{1.9e-4}\\
				8 &  \num{2.0e-4}\\
				9 &  \num{3.0e-5}\\
				\hline
			\end{tabular}
		\end{minipage}%
		\begin{minipage}{0.3\columnwidth}
			\begin{tabular}{|>{\centering\arraybackslash}p{1cm}|>{\centering\arraybackslash}p{2cm}|}
				\hline
				$\bm{n}$ & $\bm{\eta}$ \\\hline\hline
				0 &  \num{9.9e-7}\\
				1 &  \num{2.9e-6}\\
				2 &  \num{7.1e-6}\\
				3 &  \num{7.8e-6}\\
				4 &  \num{2.1e-5}\\
				5 &  \num{3.9e-5}\\
				6 &  \num{6.3e-5}\\
				7 &  \num{2.3e-4}\\
				8 &  \num{3.9e-4}\\
				9 &  \num{4.3e-4}\\
				\hline
			\end{tabular}
		\end{minipage}%
		\begin{minipage}{0.3\columnwidth}
			\begin{tabular}{|>{\centering\arraybackslash}p{1cm}|>{\centering\arraybackslash}p{2cm}|}
				\hline
				$\bm{n}$ & $\bm{\eta}$ \\\hline\hline
				0 &  \num{8.7e-7}\\
				1 &  \num{2.6e-6}\\
				2 &  \num{4.6e-6}\\
				3 &  \num{1.2e-5}\\
				4 &  \num{1.8e-5}\\
				5 &  \num{3.7e-5}\\
				6 &  \num{8.1e-5}\\
				7 &  \num{2.1e-4}\\
				8 &  \num{5.0e-4}\\
				9 &  \num{6.5e-4}\\
				\hline
			\end{tabular}
		\end{minipage}%
		\caption{\small Value of $\eta$ for $r_- = 0$ (left), $r_- = 0.25$ (middle) and $r_- = 0.5$ (right). We have fixed $r_+ = 1$, $\lambda = 2$, $N = 1000$. We solved the equation \eqref{eq:eq-det-BCL-inverted} for $m=0$.}
		\label{tab:check-YO}
	\end{table}
}

\section{Discussion and conclusions}
In the present work, we have implemented a numerical computation of the QNM frequencies based on the first-order form of the BH perturbation equations, using a matrix continued fraction method. To apply this method, an important ingredient is the knowledge of the asymptotic behaviours of the modes near the BH horizon and at spatial infinity. This information is obtained by using our previous works which provided an algorithm that can systematically identify the asymptotics of such first-order systems.

Here, we have applied our method to the Schwarzschild case, using the first-order form of the GR perturbed equations rather than the second-order Schr\"odinger equation, as a way to check the validity of this new approach, qualitatively and quantitatively. We have indeed verified that the Schwarzschild QNMs can be obtained in this way with a high precision. 

We have then used the same method to compute, for the first time,  the QNM frequencies of an exact BH solution obtained in a particular Horndeski theory. Interestingly, since the BH solution depends on a parameter that quantifies the deviation from the Schwarzschild solution, we can visualise the migration, in the complex plane,  of the initial Schwarzschild QNMs as this parameter is increased. To check the robustness of our numerical results, we have performed various tests and compare our results with those based on other techniques.

For simplicity, we have restricted the present work to axial perturbations, where the scalar field perturbation vanishes, so that there is only one degree of freedom as in GR, even if the equations of motion are different. In this case, the recurrence relation involves matrices defined in a 2-dimensional vector space. 
Since this system can be easily rewritten in a second-order  Schrödinger-like equation, we were able to check the accuracy of our numerical method versus existing methods that rely on such a reformulation. This constitutes a proof of concept for our numerical method, confirming that it gives high-precision results.
	
Moreover, our method should be readily  applicable to a first-order system of any dimension. In particular, we expect the same technique to be applicable to polar perturbations, which contain the scalar field perturbation in addition to one gravitational mode, leading to a  recurrence relation  defined in a 4-dimensional space. We plan to pursue this in future work.

Beyond the specific numerical techniques used to compute the quasi-normal of a specific solutions, let us mention that  recent discussions have raised interesting questions concerning  the so-called spectral instability~\cite{Jaramillo:2020tuu,Jaramillo:2021tmt,Destounis:2021lum,Konoplya:2022pbc}. Indeed, it has been observed that, under some hypotheses, a ``small-scale'' deviation from General Relativity could lead to a ``large'' deviation of the overtones, not necessary of very high (complex) frequency. These observations question the relevance of using QNM to test deviations from GR in the ringdown, even though the deviation is very ``small''. However, in the case of the BCL Black Hole, we do not see such an instability: overtones  deviate from the ones of Schwarzschild but in a linear controlled  manner. Hence, it would be interesting to understand why some deviations lead indeed to a spectral instability and why others do not.

\acknowledgments
 This work was supported by the
French National Research Agency (ANR) via Grant No. ANR-22-CE31-0015-01 associated with the project StronG. HR thanks Thibault Damour and François Larrouturou for interesting discussions about the mathematical aspects of the continued fraction method and BH perturbation theory. KN is very grateful to Jose-Luis Jaramillo  for his explanations on the  spectral instability of QNM.

\appendix

\section{recurrence  relation for the BCL black hole}
\label{app:rec-rel-BCL}

The matrix coefficients entering  the 5-term recurrence relation  \eqref{eq:rec-rel-BCL}  that we have obtained for the BCL black hole are given by
\begin{align}
	\alpha_n &= \begin{pmatrix} \frac{n+1-i r_0 \omega}{r_+} & i\omega \\ i r_+ (2r_-+r_+)\omega & (r_++r_-)^2 \frac{n+1-i r_0 \omega}{r_+} \end{pmatrix} \,, \quad
	\beta_n = \begin{pmatrix} \frac{-2n-1+i\omega(2r_++2r_0-r_-)}{r_+} & -\frac{2i\lambda(r_++r_-)}{r_+^3 \omega} \\ -4ir_+ r_- \omega &\beta_{22}\end{pmatrix} \,,\nonumber\\
	\gamma_n &= \begin{pmatrix} \frac{n - i\omega(r_+ + r_0 - r_-)}{r_+} & \frac{2i\lambda(2r_++3r_-)}{r_+^3\omega} \\ 2ir_+r_-\omega &\gamma_{22}\end{pmatrix} \,, \quad 
	\delta_n = \begin{pmatrix} 0  &-\frac{2i\lambda(r_+ + 3r_-)}{r_+^3\omega} \\ 0 & \delta_{22}\end{pmatrix} \,, \nonumber\\
	\varepsilon_n & = \begin{pmatrix} 0  & \frac{2i\lambda r_-}{r_+^3\omega} \\ 0 & \frac{r_-^2}{r_+} \qty[n-3 -i\omega(r_+ + r_0 - r_-)] \end{pmatrix} \,.\nonumber
\end{align}
The expressions of the coefficients $\beta_{22}$, $\gamma_{22}$ and $\delta_{22}$ are given by
\begin{eqnarray*}
\beta_{22} & = &  \frac{r_++r_-}{r_+} \qty[-2n(2r_-+r_+) + r_+ + i\omega(2r_+^2 + r_+(2r_0 + r_-) + r_-(4r_0-r_-))]  ,\\
\gamma_{22} & = & \frac{3i r_-^3 \omega}{r_+} + \frac{r_-^2}{r_+} \qty[6(n-1) - i\omega(r_+ + 6r_0)] + r_- \qty[6n - 8 - i\omega(5r_+ + 6r_0)] + r_+ \qty[n-2 -i\omega(r_+ + r_0)] ,\\
\delta_{22} & = & \frac{r_-}{r_+} \qty[-2n(r_+ + 2r_-) + 5r_+ + 8r_- + i\omega(-3r_-^2 + 2r_+(r_0+r_+) + 2r_-(2r_0 + r_+))]  \, .
\end{eqnarray*}

One recovers the coefficients for the Schwarzschild recurrence  relation given in \eqref{eq:coeffs-schwa-rec-rel} in the limit $r_+ = \mu$ and $r_- = 0$.

\section{Gaussian reduction for a 5-term recurrence  relation}
\label{app:gauss-reduc-BCL}

In this appendix, we describe the Gaussian reduction procedure that enables us to reduce the five-term recurrence relation \eqref{eq:rec-rel-BCL} into a three-term recurrence relation of the form~\eqref{eq:gauss-red-induction}. We give a proof by induction. 
Let us assume that there exists $p \geq 3$ such that  the five-term recurrence relation  \eqref{eq:rec-rel-BCL}  for all $n \leq p$ can be equivalently reformulated into the form
\begin{equation}
	\tilde{\alpha}_n Y_{n+1} + \tilde{\beta}_n Y_n+ \tilde{\gamma}_n Y_{n-1} = 0 \,,
	\label{eq:3term-rec-i}
\end{equation}
for all $n \leq p$  as well. Then, we want to prove that it is possible to cast the next order,
\begin{equation}
	\alpha_{p+1} Y_{p+2} + \beta_{p+1} Y_{p+1} + \gamma_{p+1} Y_{p} + \delta_{p+1} Y_{p-1} + \varepsilon_{p+1} Y_{p-2} = 0 \,,
	\label{eq:5term-rec-n+1}
\end{equation}
into a similar form.

The first step consists in using  \eqref{eq:3term-rec-i} at order $p-1$,
\begin{equation}
	\tilde{\alpha}_{p-1} Y_{p} + \tilde{\beta}_{p-1} Y_{p-1} + \tilde{\gamma}_{p-1} Y_{p-2} = 0 \,.
\end{equation}
If we assume  $\tilde{\gamma}_{p-1}$ to be invertible, we can eliminate $Y_{p-2}$ in  \eqref{eq:5term-rec-n+1} and we obtain,
\begin{equation}
	\alpha_{p+1} Y_{p+2} + \beta_{p+1} Y_{p+1} + \big(\gamma_{p+1} - \varepsilon_{p+1}\cdot \tilde{\gamma}_{p-1}^{-1} \cdot \tilde{\alpha}_{p-1} \big) Y_{p} + \big(\delta_{p+1} - \varepsilon_{p+1} \cdot \tilde{\gamma}_{p-1}^{-1}\cdot \tilde{\beta}_{p-1} \big) Y_{p-1} = 0 \,.
\end{equation}
Now, we  use once more \eqref{eq:3term-rec-i} but now at order $p$ so that we can express $Y_{p-1}$ in terms of $Y_p$ and $Y_{p+1}$ assuming that $ \tilde{\gamma}_{p} $ is invertible. Hence, after a direct calculation, we end up with the 3-term recurrence relation,
\begin{multline}
	\alpha_{p+1} Y_{p+2} + \Big[\beta_{p+1} - \big(\delta_{p+1} - \varepsilon_{p+1} \cdot \tilde{\gamma}_{p-1}^{-1}  \cdot \tilde{\beta}_{p-1} \big)  \cdot \tilde{\gamma}_{p}^{-1}  \cdot\tilde{\alpha}_p\Big] Y_{p+1} \\
	+ \Big[\big(\gamma_{p+1} - \varepsilon_{p+1}  \cdot \tilde{\gamma}_{p-1}^{-1}  \cdot \tilde{\alpha}_{p-1} \big) - \big(\delta_{p+1} - \varepsilon_{p+1}  \cdot \tilde{\gamma}_{p-1}^{-1}  \cdot \tilde{\beta}_{p-1} \big) \cdot  \tilde{\gamma}_{p}^{-1}  \cdot \tilde{\beta}_p \Big]Y_{p} = 0 \,.
\end{multline}
Finally, as the hypothesis  \eqref{eq:3term-rec-i} is true for any $n \leq 3$ (i.e. for $p=3$), the five-term recurrence relation can be equivalently reformulated as \eqref{eq:3term-rec-i} for any $n$. Furthermore, the new matrix coefficients  $\tilde{\alpha}_{n+1}$, $\tilde{\beta}_{n+1}$ and $\tilde{\gamma}_{n+1}$ can be computed recursively from
\begin{align}
	\tilde{\alpha}_{n+1} &= \alpha_{n+1} \,,\\
	\tilde{\beta}_{n+1} &= \beta_{n+1} - \big(\delta_{n+1} - \varepsilon_{n+1} \cdot \tilde{\gamma}_{n-1}^{-1}  \cdot \tilde{\beta}_{n-1} \big)  \cdot  \tilde{\gamma}_{n}^{-1}  \cdot  \tilde{\alpha}_n \,,\\
	\tilde{\gamma}_{n+1} &= \big(\gamma_{n+1} - \varepsilon_{n+1}  \cdot  \tilde{\gamma}_{n-1}^{-1}  \cdot  \tilde{\alpha}_{n-1} \big) - \big(\delta_{n+1} - \varepsilon_{n+1}  \cdot \tilde{\gamma}_{n-1}^{-1}  \cdot  \tilde{\beta}_{n-1} \big)  \cdot  \tilde{\gamma}_{n}^{-1}  \cdot  \tilde{\beta}_n \,.
\end{align}

\section{Comparison with other methods}
\label{app:other-methods}

In order to further verify the authenticity of the computed QNMs, it is important to compare their value outside of the Schwarzschild limit $r_- = 0$. However, until now, there has been no investigation of the QNMs of the BCL black hole in the literature. Therefore, we adapt existing methods to the case of this black hole and compare the results obtained.

In the following, we use the fact that axial perturbations of the BCL black hole can be cast into the Schrödinger-like form \cite{Langlois:2021aji}
\begin{equation}
	-\dv[2]{\Psi}{r_*} + (V(r) - \omega^2)\Psi = 0 \,,
	\label{eq:schrodinger-equation}
\end{equation}
with the potential given by
\begin{equation}
	V(r) = \frac{(r-r_+)(r+r_-)}{r^4 (r^2 + 2 r_+ r_-)^3} \times \sum_{k = 0}^6 p_k r^k \,,
	\label{eq:potential-BCL}
\end{equation}
where the coefficients $p_k$ are given by
\begin{align}
	p_0 &= r_+^3 r_-^3 (16 \lambda - 5) \,, & p_1 &= 3 r_+^2 r_-^2 (r_- - r_+) \,, & p_2 &= 6 r_+^2 r_-^2 (4\lambda - 1) \,,\nonumber\\
	p_3 &= 4 r_+ r_- (r_- - r_+) \,, & p_4 &= 3 r_+ r_- (4\lambda - 1) \,, & p_5 &= 3 (r_- - r_+)\,,\nonumber\\
	p_6 &= 2(\lambda + 1) \,.
\end{align}
The tortoise coordinate $r_*$ is defined such that
\begin{equation}
	\dv{r_*}{r} = \frac{1}{A(r)} \,,
\end{equation}
with $A(r)$ given in \eqref{eq:A-B-C-BCL}.

\subsubsection{WKB method}

The WKB method allows one to compute QNMs by approximating the potential $V$ near its maximum. It was proposed originally in~\cite{Schutz:1985km}, improved in~\cite{Iyer:1986np, Iyer:1986nq} and further improved in~\cite{Konoplya:2003ii,Matyjasek:2017psv} (see~\cite{Konoplya:2019hlu} for a review). The results of~\cite{Iyer:1986nq}, sufficient for our computation, can be summed up as follows: if the potential $V$ seen as a function of $r_*$ has a maximum at $r_* = r_{*0}$, then one can decompose the potential as
\begin{equation}
	V(r_*) = V_0 + \frac12 V_0^{(2)} (r_* - r_{*0})^2 + ... = V_0 + \sum_{k=2}^{+\infty} \frac{1}{k!} V_0^{(k)} (r_* - r_{*0})^k \,.
	\label{eq:WKB-taylor-potential}
\end{equation}
If one truncates the decomposition of~\eqref{eq:WKB-taylor-potential} at order 6, then the QNMs $\omega_n$ are given by the solutions of 
\begin{equation}
	\omega_n^2 = \qty(V_0 + \sqrt{-2 V_0^{(2)}} \Lambda) - i \qty(n + \frac12) \sqrt{-2 V_0^{(2)}} (1 + \Omega) \,,
	\label{eq:WKB-equation}
\end{equation}
where $\Lambda$ and $\Omega$ are functions of $V_0^{(2)}, ..., V_0^{(6)}$ whose expressions are given in~\cite{Iyer:1986nq}.

Equation~\eqref{eq:WKB-equation} can then be solved for $\omega_n$. In table~\ref{tab:comparison-WKB-contfrac}, we compare the QNM values obtained via the matrix continued fraction method and the WKB method for a nonzero $r_-$.

\begin{table}[!htb]
	\centering
	\begin{tabular}{|c|c|c|}
		\hline
		$\bm{n}$ & \textbf{Matrix continued fraction} & \textbf{WKB method} \\\hline\hline
		0 & 0.843719 - 0.197227i & 0.843204 - 0.19805i\\
		1 & 0.794262 - 0.607533i & 0.793898 - 0.60744i\\
		2 & 0.720858 - 1.059862i & 0.715865 - 1.03543i\\
		3 & 0.661666 - 1.556088i & 0.615373 - 1.47301i\\
		4 & 0.632640 - 2.077741i & 0.490639 - 1.91668i\\
		5 & 0.627338 - 2.610898i & 0.340383 - 2.36668i\\
		6 & 0.636227 - 3.149321i & 0.164472 - 2.82404i\\
		\hline
	\end{tabular}
	\caption{\small Comparison between the first QNM values obtained via our numerical method and the WKB approximation, for $\lambda = 2$, $r_+ = 1$, $r_- = 0.5$. }
	\label{tab:comparison-WKB-contfrac}
\end{table}
We observe that the first overtones are very well approximated by the WKB method while this method fails at higher values of $n$.

\subsubsection{Spectral decomposition}

We can also compare our QNM computation with another fully numerical computation, based on the numerical resolution of the Schrödinger-like reformulation of the axial perturbations equations. This method was developed in~\cite{Jansen:2017oag} and relies on a spectral decomposition of the mode function in order to cast the QNM computation problem into a generalized eigenvalue problem. It is available as a Mathematica package and only requires the Schrödinger-like equation~\eqref{eq:schrodinger-equation} and the potential~\eqref{eq:potential-BCL}, rescaled in an appropriate way. 

In table~\ref{tab:comparison-WKB-QNMSpectral}, we compare the QNM values obtained via the matrix continued fraction method and this numerical method (called \texttt{QNMSpectral}) for a non-zero $r_-$.

\begin{table}[!htb]
	\centering
	\begin{tabular}{|c|c|c|}
		\hline
		$\bm{n}$ & \textbf{Matrix continued fraction} & \textbf{\texttt{QNMSpectral} method} \\\hline\hline
		0 & 0.84371827443 - 0.19722719264i & 0.84371827438 - 0.19722719255i\\
		1 & 0.79426238123 - 0.60753311430i & 0.79426237852 - 0.60753340736i\\
		2 & 0.72085805503 - 1.05986135470i & 0.72088369854 - 1.05978730679i\\
		\hline
	\end{tabular}
	\caption{\small Comparison between the first QNM values obtained via our numerical method and the \texttt{QNMSpectral} numerical routine, for $\lambda = 2$, $r_+ = 1$, $r_- = 0.5$.}
	\label{tab:comparison-WKB-QNMSpectral}
\end{table}
 We observe that both methods agree up to a very high precision on the first overtones (we used a convergence threshold of $10^{-14}$ to account for this). However, it is not possible to reach higher values of $n$ using the \texttt{QNMSpectral} approach (and this method requires a Schrödinger-like reformulation, which is not available for polar perturbations).

\bibliographystyle{utphys}
\bibliography{full_biblio}

\end{document}